\documentclass[conference]{IEEEtran}
\ifCLASSINFOpdf
\else
\fi
\usepackage{fancyhdr}
\usepackage{graphicx} 
\usepackage{epsfig}
\usepackage{multirow}
\usepackage[center]{caption}
\usepackage{subcaption}
\usepackage{setspace}
\usepackage{amsmath}
\usepackage{float}
\usepackage{cite}
\usepackage{array}
\usepackage{verbatim}
\usepackage{graphicx} 
\usepackage{amsfonts}
\usepackage{amssymb}
\onecolumn
\begin{document}
\title{Computationally efficient MIMO system identification using Signal Matched Synthesis Filter Bank}
\author{Binish~Fatimah and
S.~D.~Joshi~\\
Department
of Electrical  Engineering, Indian Institute of Technology, New Delhi, India  110016 \\
e-mail: binish.fatimah@ee.iitd.ac.in, sdjoshi@ee.iitd.ac.in

}

\markboth{}%
{}
\maketitle

\begin{abstract}
We propose a multi input multi output(MIMO) system identification framework by interpreting the MIMO system in terms of a multirate synthesis filter bank. The proposed methodology is discussed in two steps: in the first step the MIMO system is interpreted as a synthesis filter bank and the second step is to convert the MIMO system into a SISO system ``without any loss of information", which re-structures the system identification problem into a SISO form. The system identification problem, in its new form, is identical to the problem of obtaining the signal matched synthesis filter bank (SMSFB) as proposed in \cite{PartII}. Since we have developed fast algorithms to obtain the filter bank coefficients in \cite{PartII}, for ``the given data case" as well as ``the given statistics case", we can use these algorithm for the MIMO system identification as well. This framework can have an adaptive as well as block processing implementation. 
The algorithms, used here, involve only scalar computations, unlike the conventional MIMO system identification algorithms where one requires matrix computations. These order recursive algorithm can also be used to obtain approximate smaller order model for large order systems without using any model order reduction algorithm. The proposed identification framework can also be used for SISO LPTV system identification and also for a SIMO or MISO system. The efficacy of the proposed scheme is validated and its performance in the presence of measurement noise is illustrated using simulation results.

\end{abstract}


\section{Introduction}
This paper addresses the problem of MIMO system identification, which is required in various applications like human control behavior, modeling of unmanned aerial vehicles, robotic assemblies, chemical plants, power systems, multiuser channel estimation in communication systems, stock market analysis etc.\cite{MIMO_commu.,MIMO_Power_SS,MIMO_OFDM,MIMO_aerial,MIMO_chemical,MIMO_human}. The objective of MIMO identification is to estimate the parameters of the system, which can either be described by a state space model \cite{MIMO_SS,MIMO_SS_1} or an FIR model \cite{MIMO_FIR}, given the input and output signals or only with the information of the output signals.

 In this paper, given the observed input and output signals, we estimate the FIR model structure of a MIMO system using a variate form of signal matched synthesis filter bank(SMSFB) as defined \cite{PartII}. An analogy is established between the MIMO system identification problem and the problem of estimating the SMSFB. Since we have developed fast algorithms to estimate the filter coefficients, in \cite{PartII}, we use the algorithm in this work for system identification. Most of the MIMO system identification methods, existing in the literature, suffer from two main drawbacks: 1) development of adaptive algorithms becomes cumbersome, and 2) parameter estimation involves matrix computations \cite{MIMO_1,MIMO_2,MIMO_3}. In contrast, the system identification algorithm used in this work involves only scalar computations and can be implemented in a block processing mode as well as in an adaptive manner. 
 
 In control theory we come across various large scale complex systems which are approximated using smaller order models for computational efficiency or for design simplification. One of the existing approaches decompose the large order system into a number of decoupled sub-systems or small scale systems. Each sub-system is individually modelled and then using a coordinator the subsystems are adjusted such that the required response is achieved. Since the order recursive algorithm used in this paper have a lattice-like implementation, they can be used to approximate large order systems using a smaller order model in a fast and computationally efficient manner. Also, the decentralized systems can be naturally modelled using the proposed scheme. 

Control theory community have used theories and results from signal processing literature for example ARMA models. It has also been discussed by authors like Ljung \cite{Ljung_1991} that a more extensive relationship between the two areas should be explored which will lead to promising results. This work is an example of a similar attempt. Here we use the results as well as algorithms existing in the multirate filter bank literature to propose computationally efficient system identification framework.
 
 This paper is organized as follows: in section II we present some preliminaries from the multirate filter bank theory and present a MIMO form of synthesis filter bank. This representation is, then, used to interpret a MIMO system in terms of a synthesis filter bank. In section III we convert a MIMO system into a SISO form and section IV discusses the MIMO system identification problem and re-structured the problem into a SISO form. The estimation problem, in this case, is identical to obtaining signal matched synthesis filter bank parameters, proposed in \cite{PartII} which can be exploited to obtain the required MIMO system. Since fast algorithms for SMSFB have already been developed in \cite{PartII,given_stat} for ``given data" as well as ``given statistics" cases, these algorithms can be used to compute the MIMO parameters in a computationally efficient manner. Thus, the MIMO parameters can be obtained if one realization of the process is available, also in real time and also for the cases where the statistics of the signal are known or can be estimated. The efficacy of the proposed algorithm is validated using simulation results in section V and conclusions are discussed in section VI.   

\subsection{Notations}
In this paper we denote random variables by italic lowercase letters, vectors of random variable by italic, boldfaced, lowercase letters and matrices of random variables with italic capital letters. The norm of a vector  $\textit{\textbf{x}}$ is denoted by  $\mid \mid \textit{\textbf{x}} \mid \mid$, which is the positive square-root of the inner-product of $\textit{\textbf{x}}$ with itself, where $\textit{\textbf{x}}$ belongs to a Hilbert space.

\section{MIMO system and multirate synthesis filter bank}
In this section we interpret a MIMO system in terms of a multirate synthesis filter bank. First we briefly discuss the theory of synthesis filter bank and some of its results.
 \subsection{Preliminaries}
In this section we briefly review some relevant results from the filter bank theory which are required for better understanding and appreciation of the results presented in this paper. An extensive treatment of the theory of multirate filter banks is available in \cite{Vaidyanathan}.
\vspace*{-1cm}
\begin{figure}[H]
\includegraphics[width=1\textwidth,height=0.5\textwidth]{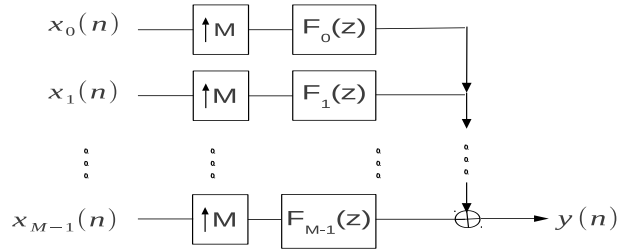}
\vspace{-3.8cm}
\caption{M-channel multirate synthesis filter bank} \label{fig:2.multirate_filter_bank}
\end{figure}
A set of digital filters represented by a multi input single output system, with each filter preceded by an upsampler, is called a synthesis filter bank, shown in Figure \ref{fig:2.multirate_filter_bank}. An M-fold upsampler increases the sampling rate by  a factor of M by appending M-1 zeros between every two samples of the input signal.

Let us consider the i-th filter, of the M-band synthesis filter bank, denoted as $F_{i}(z)$. Assuming the filter to be FIR and filter order to be multiple of M, without any loss of generality, we can write :
\begin{eqnarray} 
&&\rm {F}_{i}(z)= \sum_{n=0}^{N-1}{f_{i}(n)z^{-n}}=\begin{small}\left(\sum_{p=0}^{(N/M-1)}{{f}_{i}(pM)z^{-Mp}}\right)
+z^{-1}\left(\sum_{p=0}^{(N/M-1)}{{f}_{i}(pM+1)z^{-Mp}}\right)\end{small}
\nonumber\\ 
&&\hspace{1cm}+\cdots+\rm  z^{-(M-1)}\left(\sum_{p=0}^{(N/M-1)}{{f}_{i}(pM+M-1)z^{-Mp}}\right)\nonumber\\
&&\hspace{1cm}\rm \equiv{ F_{i,M-1}(z^M)+z^{-1}F_{i,M-2}(z^M)}+
 \cdots + z^{-(M-1)}F_{i,0}(z^M),\hspace{0.5cm} 0 \leq i\leq M-1.\nonumber
\end{eqnarray}

The M-components, written in the braces, are called the type-II polyphase components. The k-th component for the i-th filter can be written as:\\
\begin{eqnarray}
\rm {F}_{i,k}(z^M) \triangleq \sum_{p=0}^{(N/M-1)}{{f}_{i}(pM+M-1-k)z^{-Mp}}, \hspace{0.5cm} 0 \leq i\leq M-1.
\end{eqnarray}

\begin{figure}[H]
\hspace{3.5cm}
\includegraphics[width=0.6\textwidth,height=0.4\textwidth]{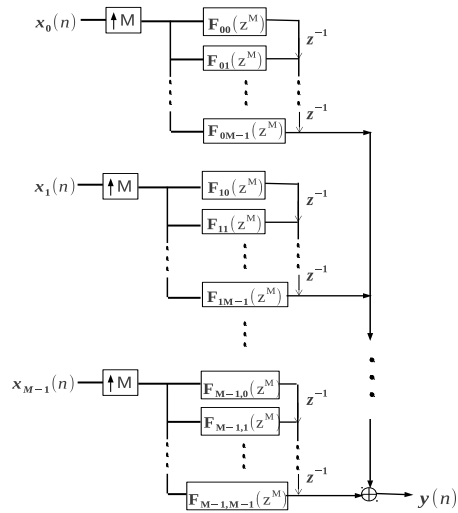}
\caption{Synthesis Filters in terms of type-II polyphase components} \label{fig:synthesis_poly_fig1}
\end{figure}

The synthesis filter bank can be represented using the polyphase components, as shown in Fig. \ref{fig:synthesis_poly_fig1}.

\vspace*{-4cm}
\begin{figure}[H]
\includegraphics[width=1\textwidth,height=0.6\textwidth]{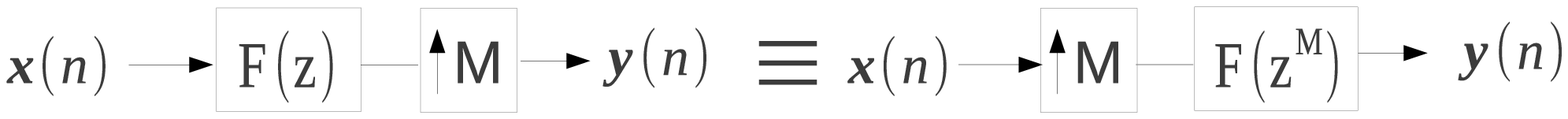}
\vspace{-5cm}
\caption{Noble Identity for Multirate systems} \label{fig:noble_identities}
\end{figure}

\begin{figure}[H]
\hspace{5cm}
\includegraphics[width=0.6\textwidth, height=0.4\textwidth]{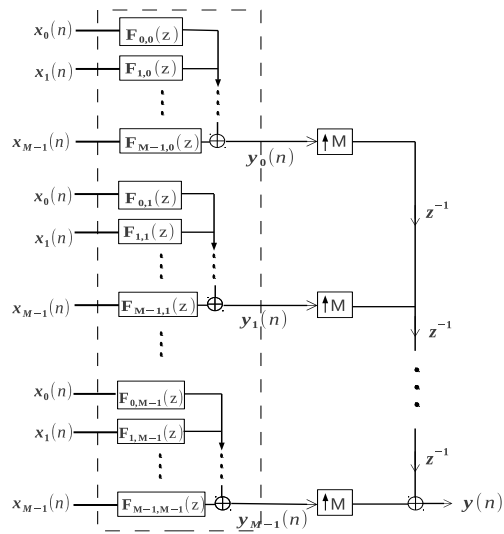}
\caption{Polyphase decomposition of Synthesis filter bank} \label{fig:4.poly}
\end{figure}

The polyphase components of synthesis filters and upsamplers, given in Fig.{\ref{fig:synthesis_poly_fig1}}, can be re-arranged using the noble identity (given in Fig.\ref{fig:noble_identities}) to obtain a computationally efficient structure for M-channel synthesis filter bank, as shown in Fig.\ref{fig:4.poly}. 



From Figure \ref{fig:4.poly} we can write the output $y(n)$ as follows:
\begin{align*}
y(n)=\left\{ \begin{array}{l l} y_{0}(n)\hspace{0.5cm}\mbox{ if }{n=iM} ,\mbox{where $i$ is an integer}\\
 y_{1}(n)\hspace{0.5cm}\mbox{ if }{n=iM-1},\\
 \vdots\\
 y_{M-1}(n)\hspace{0.25cm}\mbox{ if }{n=iM-M+1} \end{array}\right.
\end{align*}
From Figure \ref{fig:4.poly} the outputs, $y_{i}(n)$, can be written in the following form (using convolution):
\begin{align}
& {{ y}}_{i}(n)= \sum_{j=0}^{M-1}{\sum_{k=0}^{N-1}{x_{j}(n-k)\rm{f}_{i,j}(k)}}, \hspace{0.5cm} 0 \leq i\leq M-1,\label{eq:4.poly}\\
 & \hspace*{0.8cm} = \sum_{j=0}^{M-1}{\sum_{k=0}^{N-1}{x_{j}(n-k)\rm{f}_{j}(Mk+M-1-i)}}, \hspace{0.5cm} 0 \leq i\leq M-1. 
\end{align}

The above set of equations, for $0 \leq i \leq M-1$, can be written in a matrix form as follows:\\

\renewcommand{\arraystretch}{1.2}
\begin{footnotesize}
\begin{eqnarray} &\hspace*{-0.5cm}
\left[\begin{array}{c}
{y}_{0}(n)\\
{y}_{1}(n)\\
 \vdots \\
{y}_{M-1}(n)
\end{array}\right] =\left[\begin{array}{cccccc}
\rm f_{0}(M-1) & \rm f_{1}(M-1) & \ldots & \rm f_{M-1}(M-1)\\
\rm f_{0}(M-2) & \rm f_{1}(M-2) & \ldots & \rm f_{M-1}(M-2)\\
\vdots & \vdots & \ldots & \vdots\\
\rm f_{0}(0) & \rm f_{1}(0) & \ldots & \rm f_{M-1}(0)
\end{array}\right]
\left[\begin{array}{c}
\textit{x}_{0}(n)\\
\textit{x}_{1}(n)\\
\vdots \\
\textit{x}_{M-1}(n)
\end{array}\right]+...+
\nonumber \\&\hspace*{0.5cm}
\left[\begin{array}{cccccc}
\rm f_{0}(N-M) & \rm  f_{1}(N-M) & \ldots & \rm f_{M-1}(N-M)\\
 \rm f_{0}(N-M-1) & \rm  f_{1}(N-M-1) & \ldots & \rm f_{M-1}(N-M-1)\\
\vdots & \vdots & \ldots & \vdots \\
 \rm f_{0}(N-1) & \rm  f_{1}(N-1) & \ldots &  \rm f_{M-1}(N-1)
\end{array}\right]
\left[\begin{array}{c}
\textit{x}_{0}(n-N+1)\\
\textit{x}_{1}(n-N+1)\\
\vdots \\
\textit{x}_{M-1}(n-N+1)
\end{array}\right].
\label{eq:4.2}
\end{eqnarray}
\end{footnotesize}\renewcommand{\arraystretch}{1}
Equation (\ref{eq:4.2}) is a multi-input multi-output (MIMO) form of the synthesis filter bank. It is pertinent to note that i-th channel output, i.e. ${y}_{i}(n)$, makes use of all the past inputs of all the input channels, i.e. $w_{i}(k)$ for $0 \leq i \leq M-1$ and  $n-N+1\leq k \leq n$. 

 With these pre-requisites at our disposal we are now in a position to discuss the interpretation of a MIMO system in terms of multirate synthesis filter bank. 

\subsection{Interpretation of MIMO system as multirate synthesis filter bank}
In this sub-section we represent a MIMO system in terms of a synthesis filter bank. This interpretation is used in the next section to propose the system identification algorithm. 
\vspace*{-1cm}
\begin{figure}[H]
\hspace{4cm}
\includegraphics[width=12cm, height=9cm]{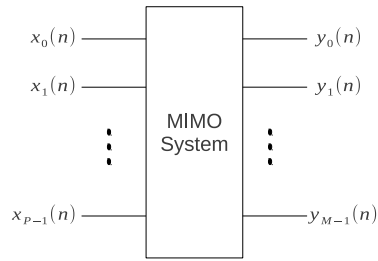}
\vspace{-3cm}
\caption{N-input M-output system} \label{fig:6.MIMO}
\end{figure}

A MIMO system with P inputs and M outputs, Fig. \ref{fig:6.MIMO}, can be represented using an FIR model given as follows:
\begin{align}
\left[\begin{array}{c}
Y_{0}(z)\\
Y_{1}(z)\\
\vdots\\
Y_{M-1}(z)
\end{array}\right]=
\left[\begin{array}{c}
 \rm H_{0}(z)\\
 \rm H_{1}(z)\\
\vdots\\
 \rm H_{M-1}(z)
\end{array}\right]
\left[\begin{array}{c}
X_{0}(z)\\
X_{1}(z)\\
\vdots\\
X_{P-1}(z)
\end{array}\right]
\end{align}
where, $Y_{i}(z)=\sum_{n=-\infty}^{\infty}{y_{i}(n)z^{-n}}$, $X_{i}(z)=\sum_{n=-\infty}^{\infty}{x_{i}(n)z^{-n}}$ and $ \rm H_{i}(z)=\sum_{n=0}^{N'-1}{h_{i}(n)z^{-n}}$(without any loss of generality we can assume N' to be a multiple of P). In time domain, the above equation can be written as: 
\begin{align}
&\left[\begin{array}{c}
 {y}_{0}(n)\\
{y}_{1}(n) \\
\vdots \\
{y}_{M-1}(n)
\end{array}
\right]=
\left[\begin{array}{cccc}
 \rm h_{0}(0) &  \rm h_{0}(1) & \cdots &  \rm h_{0}(P-1)\\
 \rm h_{1}(0) &  \rm h_{1}(1) & \cdots &  \rm h_{1}(P-1)\\
\vdots & \vdots & \cdots & \vdots \\
 \rm h_{M-1}(0) &  \rm h_{M-1}(1) & \cdots &  \rm h_{M-1}(P-1)
\end{array}
\right]
\left[\begin{array}{c}
x_{0}(n)\\
x_{1}(n)\\
\vdots \\
x_{P-1}(n)
\end{array}
\right]+\nonumber\\
&\cdots +
\left[\begin{array}{cccc}
 \rm h_{0}(N'-P) &  \rm h_{0}(N'-P-1) & \cdots &  \rm h_{0}(N'-1)\\
 \rm h_{1}(N'-P) & \rm  h_{1}(N'-P-1) & \cdots &  \rm h_{1}(N'-1)\\
\vdots & \vdots & \cdots & \vdots \\
 \rm h_{M-1}(N'-P) & \rm  h_{M-1}(N'-P-1) & \cdots &  \rm h_{M-1}(N'-1)
\end{array}
\right]
\left[\begin{array}{c}
x_{0}(n-N'+1)\\
x_{1}(n-N'+1)\\
\vdots \\
x_{P-1}(n-N'+1)
\end{array}
\right]
\label{eq:6.11}
\end{align}

Comparing (\ref{eq:4.2}) and (\ref{eq:6.11}), 
 we observe that the MIMO system, can be interpreted using a synthesis filter bank where the synthesis filter coefficients and the MIMO system parameters are related as:
\begin{align}
\rm H_{i}(z)=\sum_{j=0}^{Q-1}{\sum_{k=0}^{N'-1}{{f}_{j}(Qk+Q-1-i)}z^{j+Qk}}
\end{align}
where Q is the LCM(P,M) and N'=N.

Using this analogy between the multirate filter bank and MIMO system the algorithms, results and tools available for filter banks can be easily used for MIMO systems and vice-verse. As mentioned in the abstract the next step in the proposed framework is to convert the MIMO system into a SISO form which will be latter used to re-define the MIMO system identification problem as a SISO form.    

\section{SISO form of a MIMO system}
In this section we convert the MIMO system into a SISO system. In order to use the existing theories of SISO directly for MIMO systems researchers diagonalize the matrix $\rm H$, resulting into M-SISO systems. However, this operation leads to loss of information which increases the error between the estimated system and the true system and these techniques can only be used when H is a square matrix. However, the methodology proposed here converts the MIMO system into a SISO form ``without any loss of information". 
The conversion can be achieved using the following steps:
\\
\\
\textbf{Step 1: Serializing the input signals using a commutator switch.}\\
We define
%
  a scalar process, $z$, which is a scalarized form of the vector process $\boldsymbol{x}(n)$, given as follows:
%
 \begin{equation}
\textit{z}(Pn-i)=\textit{x}_i(n), \hspace{1cm} 0 \leq i\leq P-1.
\end{equation} 

\begin{figure}[H]
\hspace{3cm}
\includegraphics[width=10cm, height=9.5cm]{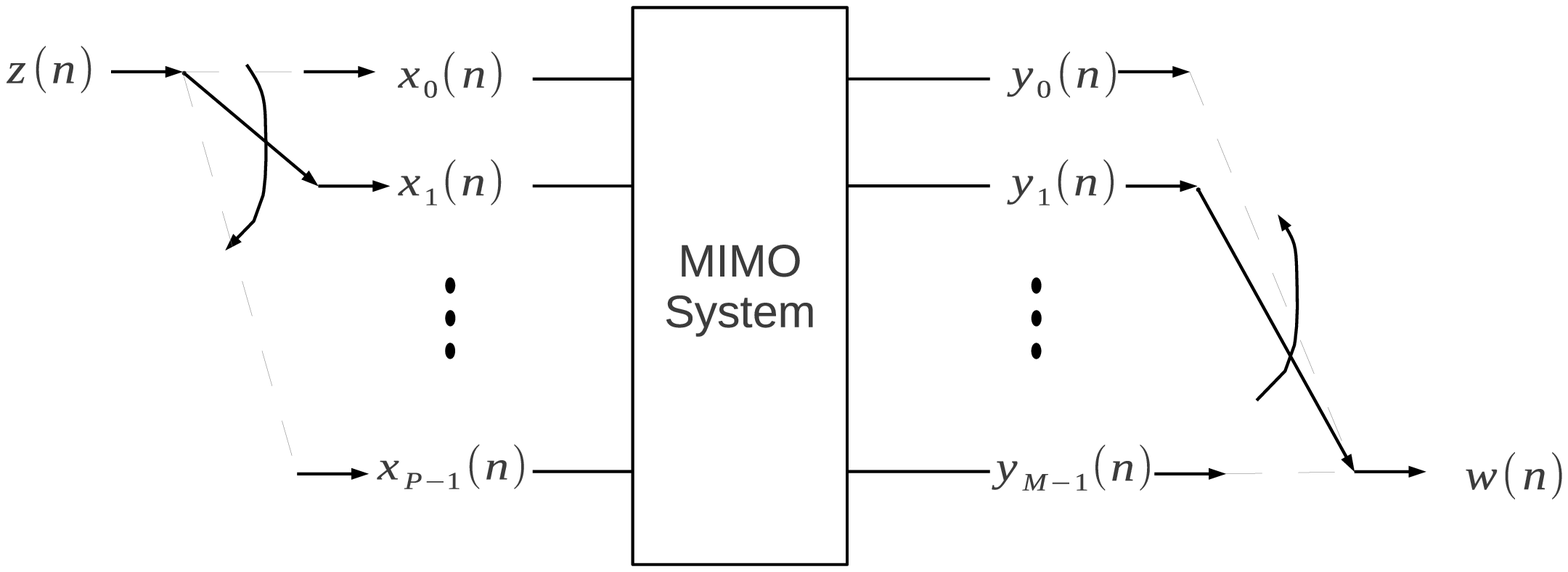}
\vspace{-100pt}
\caption{SISO form of the MIMO system} \label{fig:6.SISO}
\end{figure}

Therefore, equation(\ref{eq:6.11}) can be written as:
\begin{align}
\boldsymbol{y}(n)= \sum_{i=0}^{N'/P-1}{\rm{ H(i)}\boldsymbol{z}(Pn-i)}
 \label{eq:6.22}
\end{align}
where\\
$\boldsymbol{z}(Pn-i)=\left[\begin{array}{cccc}
z(Pn-i)&
z(Pn-1-i)&
\cdots &
z(Pn-P+1-i)
\end{array}
\right]^T$,
\\
\\
$ \rm H(i)=\left[\begin{array}{cccc}
 \rm h_{0}(P*i) & \rm  h_{0}(P*i+1) & \cdots &  \rm h_{0}(P*i+P-1)\\
 \rm h_{1}(P*i) &  \rm h_{1}(P*i+1) & \cdots &  \rm h_{1}(P*i+P-1)\\
\vdots & \vdots & \cdots & \vdots \\
 \rm h_{M-1}(P*i) &  \rm h_{M-1}(P*i+1) & \cdots &  \rm h_{M-1}(P*i+P-1)
\end{array}
\right]$ and
\\
\\
$\boldsymbol{y}(n)=\left[\begin{array}{cccc}
y_{0}(n)&
y_{1}(n)&
\cdots &
y_{M-1}(n)
\end{array}
\right]^T$.
\\
\\
\textbf{Step 2: Serializing the outputs using a commutator switch.} \\
Using a commutator switch, the M-outputs of the single input multi output(SIMO) system, given in (\ref{eq:6.22}), are interleaved to give a single output, shown in Fig.\ref{fig:6.SISO}, given as follows:
\begin{align}
{\textit{w}}(Mn-i)={\textit{{y}}}_{i}(n), \hspace{1cm} 0 \leq i\leq M-1. \label{eq:6.4.3}
\end{align}

Therefore, we can write:
\begin{eqnarray}
\boldsymbol{w}(Mn)=\sum_{i=0}^{N'/P-1}{\rm{ H(i)}\boldsymbol{z}(Pn-i)}
 \label{eq:6.22(a)}
\end{eqnarray}
where,\\
$\boldsymbol{w}(Mn)=\left[\begin{array}{cccc}
{w}(Mn)&
{w}(Mn-1) &
\cdots &
{w}(Mn-M+1)
\end{array}
\right]^T$

\vspace*{-2cm}
\begin{figure}[H]
\vspace*{-1.5cm}
\hspace{3cm}
\includegraphics[width=0.6\textwidth, height=0.4\textwidth]{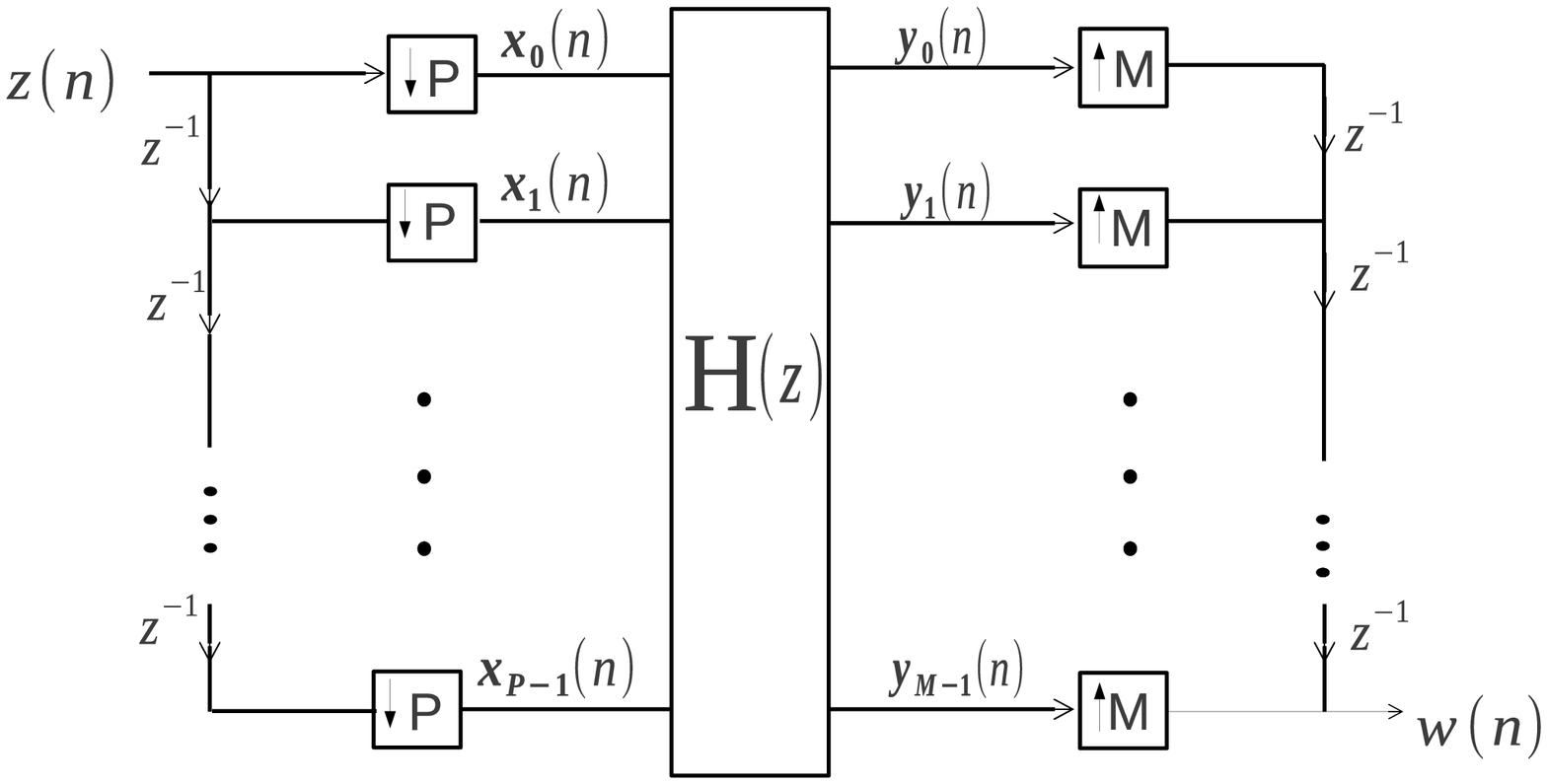}
\vspace{-40pt}
\caption{SISO interpretation of the MIMO system} \label{fig:conf.poly_synth}
\end{figure}

The MIMO system of Fig.\ref{fig:6.MIMO} can also be represented as Fig.\ref{fig:conf.poly_synth}.

Equation (\ref{eq:6.22(a)}) represents a single input single output system which is equivalent to the MIMO system represented by equation(\ref{eq:6.11}). In the next section we first present the MIMO system identification problem, then using the equivalent SISO form, given by (\ref{eq:6.22(a)}), re-frame the problem. 

\section{MIMO system identification}
\begin{figure}[H]
\begin{center}
\vspace{-20pt}
\hspace{-1cm}
\includegraphics[width=10cm, height=6cm]{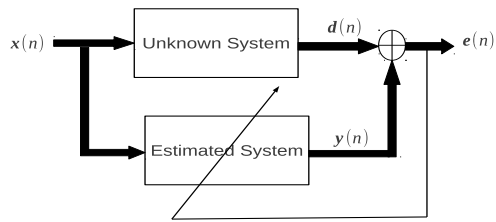}
\vspace{-40pt}
\caption{MIMO system identification block diagram} \label{fig:6.MIMO_sys}
\end{center}\end{figure}

System identification refers to the operation of mathematical modeling of dynamic systems from the measured input and output data, as shown in Figure \ref{fig:6.MIMO_sys}. 

\begin{eqnarray}
&&\boldsymbol{e}(n)= \boldsymbol{d}(n)-\boldsymbol{y}(n)\\
&&\hspace{1cm}=\boldsymbol{d}(n)-H\boldsymbol{x}(n)
, \label{eq:6.33}
\end{eqnarray}
 
 where the error vector is defined as:
 
 $\boldsymbol{e}(n)=\left[\begin{array}{cccc}
e_{0}(n)&
e_{1}(n)&
\cdots &
e_{M-1}(n)
\end{array}
\right]^T$ and \\
the measured output vector is given as:\\
 $\boldsymbol{y}(n)=\left[\begin{array}{cccc}
y_{0}(n)&
y_{1}(n)&
\cdots &
y_{M-1}(n)
\end{array}
\right]^T$.

 The system identification problem considered in this work, for an N-input and M-output system (Figure  \ref{fig:6.MIMO_sys}), can be stated as: ``given the P-inputs of the system, $x_{i}(n)$ for $ 0 \leq i \leq P-1$, and the M-outputs, $d_{i}(n)$ for $ 0 \leq i \leq M-1$, obtain the system parameters, $h_{i}(n)$, such that distance between the response of the estimated system, ${y}_{i}(n)$ and the true system, $d_{i}(n)$, for $0 \leq i \leq M-1$, is minimized". For dynamic systems, this is an adaptive identification problem. We are considering here a noiseless condition, however, the system can also be analyzed in the presence of noise.

 Using the equivalent SISO form of the MIMO system, (\ref{eq:6.22(a)}), as obtained in the previous section, the above equation can be easily written as:


\begin{eqnarray}
\boldsymbol{e}(n)=\boldsymbol{d}(n)-\rm{H}\boldsymbol{z}(Pn)
 \label{eq:6.33}
\end{eqnarray}
 
%

Here, H can be obtained in the least squares sense, as given below:
\begin{align}
\rm{H}=\boldsymbol{d}(n)\boldsymbol{z}^{T}(Mn)\left[\boldsymbol{z}(Mn)\boldsymbol{z}^{T}(Mn)\right]^{-1}\label{eq:LS_sol}
\end{align}
where, $\boldsymbol{z}^{T}(Mn)$ denotes the transpose of vector $\boldsymbol{z}(Mn)$. Objective of the given system identification problem, therefore, is the solution of the above equation. However, the solution will involve matrix inversion and its complexity will increase as the number of inputs and outputs or the filter order increases. As mentioned in the signal processing literature, a recursive solution of equation (\ref{eq:LS_sol}) will not involve any matrix computations; thus yielding a computationally efficient and fast solution.

It can be easily observed that the MIMO system identification problem, as represented by (\ref{eq:6.33}), is structurally similar to the estimation problem proposed in \cite{PartII}, for the signal matched synthesis filter bank(SMSFB). The concept of SMSFB, as proposed in \cite{PartII}, can be easily re-worked to obtain the MIMO system parameters.

In \cite{PartII} we have also developed a fast, order and time recursive least squares algorithm to compute the optimum filter parameters for the ``given data case". Moreover, for application where the statistics of the signal are available or can be computed, an order recursive Levinson-type algorithm has been developed in \cite{given_stat}, to obtain the optimum SMSFB coefficients in the mean square sense. Due to the similarity between the proposed MIMO system and SMSFB, as discussed above, the fast algorithms proposed in \cite{PartII, given_stat} can be used to obtain the MIMO system parameters. The recursions, obtained in the algorithms, gives rise to lattice-like structures. Hence, we obtain a framework for MIMO system identification which involves only scalar computations and can be implemented in both block processing mode as well as adaptive mode. 
\\

\textbf{Remarks:}
\begin{enumerate}
\item In the proposed scheme we converted the given MIMO system into an equivalent SISO system and obtained the parameters using only scalar computations. It is easy to observe that the proposed framework can be used for a SIMO system as well as a MISO system identification problem.
\item The ``given data" as well as ``given statistics" algorithms developed in \cite{PartII} and [] are order recursive algorithms and can be implemented using lattice-like structures. Therefore large order systems can be estimated using smaller order models without using any existing model reduction algorithms.  
\item  Researchers have proposed that a SISO systems with linear periodically time varying property can be implemented with multiple filters and thus can be converted to MIMO systems \cite{MIMO_LPTV}. Thus, the LPTV SISO system parameters can be obtained using the MIMO system identification proposed above.
\end{enumerate}


\section{Simulations}
In this section, the efficacy of the MIMO system identification framework, as discussed in the previous section, is validated using simulation results. The following experiment is designed for this purpose.

Experiment: For the purpose of simulation studies, we generate a signals using two AR processes, with poles at $0.5e^{-j\pi/{3}}$ and $0.9e^{-j\pi/{3}}$ respectively, and a $2 \times 2$ system, with filter order as 5. Now, given the input signals and the generated output signals the algorithm, proposed in this paper, is used to estimate the MIMO system parameters. And as mentioned earlier, the scheme proposed in this work can be analyzed in the presence of noise. For this purpose, we estimate the MIMO system parameters in the presence of measurement noise, which is white and independent to the input signals. The parameters computed for different SNR values are given in Table \ref{table:SNR_1}.
\renewcommand{\arraystretch}{1.6}
\begin{table}[H]
\begin{small}
\caption{Parameter values for different SNR} \label{table:SNR_1}
\begin{center}\begin{tabular}{|llllll|}
\hline
SNR & $h_{0}(0)$& $h_{0}(1)$ & $h_{0}(2)$ & $h_{0}(3)$  & $h_{0}(4)$ \tabularnewline
\hline
actual parameters & 1.0183 &  -1.9500  &  2.2364  & -1.7274  &  0.6192 \tabularnewline
1 &   1.0218 &  -1.9626  &  2.2733  & -1.7285   & 0.6547    \tabularnewline
0.1 & 0.9988  & -1.9238  &  2.2806  & -1.7740  &  0.6843  \tabularnewline
0.001 &  0.9903 &  -1.9183 &   2.2690 &  -1.7709  &  0.6918\tabularnewline
\hline
SNR & $h_{1}(0)$& $h_{1}(1)$ & $h_{1}(2)$ & $h_{1}(3)$  & $h_{1}(4)$ \tabularnewline
\hline
actual parameters & 0.9872  & -2.3314  &  3.4243 &  -3.0962   & 1.7222\tabularnewline
1 & 0.9453  & -2.3106  &  3.4250   & -3.1156  &  1.7222\tabularnewline
0.1 & 0.9062  & -2.3276  &  3.4055 &  -3.1027  &  1.7338 \tabularnewline
0.001 & 1.0030 &  -2.3336  &  3.4322 &  -3.0888 &   1.7059\tabularnewline
\hline
\end{tabular}
\end{center}
\end{small}
\end{table}

\section{Conclusions} 
In this work, a given MIMO system has been converted to a SISO system ``without any loss of information". This methodology can be used in various applications involving MIMO systems, for example in the design of MIMO control systems.  We have also proposed that a MIMO system can be interpreted as the synthesis side of a filter bank, which helps to establish that the MIMO system identification problem is structurally identical to the parameter estimation problem of Signal Matched Synthesis Filter Bank \cite{PartII}. This analogy is exploited to obtain a computationally efficient solution for system identification.
\bibliographystyle{IEEEtran} 
\bibliography{IEEEabrv,reference}

\begin{thebibliography}{10}
\providecommand{\url}[1]{#1}
\csname url@samestyle\endcsname
\providecommand{\newblock}{\relax}
\providecommand{\bibinfo}[2]{#2}
\providecommand{\BIBentrySTDinterwordspacing}{\spaceskip=0pt\relax}
\providecommand{\BIBentryALTinterwordstretchfactor}{4}
\providecommand{\BIBentryALTinterwordspacing}{\spaceskip=\fontdimen2\font plus
\BIBentryALTinterwordstretchfactor\fontdimen3\font minus
  \fontdimen4\font\relax}
\providecommand{\BIBforeignlanguage}[2]{{%
\expandafter\ifx\csname l@#1\endcsname\relax
\typeout{** WARNING: IEEEtran.bst: No hyphenation pattern has been}%
\typeout{** loaded for the language `#1'. Using the pattern for}%
\typeout{** the default language instead.}%
\else
\language=\csname l@#1\endcsname
\fi
#2}}
\providecommand{\BIBdecl}{\relax}
\BIBdecl

\bibitem{PartII}
\BIBentryALTinterwordspacing
B.~Fatimah and S.~D. Joshi, ``Exact least squares algorithm for signal matched
  synthesis filter bank: Part {II},'' \emph{CoRR}, vol. abs/1409.5099, 2014.
  [Online]. Available: \url{http://arxiv.org/abs/1409.5099}
\BIBentrySTDinterwordspacing

\bibitem{MIMO_commu.}
Y.-H. Kim and S.~Shamsunder, ``Multi-stage receiver for multipath and
  interference suppression in multi-user cdma,'' in \emph{Personal Wireless
  Communications, 1997 IEEE International Conference on}, Dec 1997, pp.
  125--129.

\bibitem{MIMO_Power_SS}
I.~Kamwa and L.~Gerin-Lajoie, ``State-space system identification-toward mimo
  models for modal analysis and optimization of bulk power systems,''
  \emph{Power Systems, IEEE Transactions on}, vol.~15, no.~1, pp. 326--335, Feb
  2000.

\bibitem{MIMO_OFDM}
Y.~Zou, O.~Raeesi, and M.~Valkama, ``Efficient estimation and compensation of
  transceiver non-reciprocity in precoded tdd multi-user mimo-ofdm systems,''
  in \emph{Vehicular Technology Conference (VTC Fall), 2014 IEEE 80th}, Sept
  2014, pp. 1--7.

\bibitem{MIMO_aerial}
A.~Kallapur, M.~Samal, V.~Puttige, S.~Anavatti, and M.~Garratt, ``A ukf-nn
  framework for system identification of small unmanned aerial vehicles,'' in
  \emph{Intelligent Transportation Systems, 2008. ITSC 2008. 11th International
  IEEE Conference on}, Oct 2008, pp. 1021--1026.

\bibitem{MIMO_chemical}
D.~Rivera, H.~Lee, H.~Mittelmann, and M.~Braun, ``High-purity distillation,''
  \emph{Control Systems, IEEE}, vol.~27, no.~5, pp. 72--89, Oct 2007.

\bibitem{MIMO_human}
S.~Kukreja, B.~Haverkamp, D.~Westwick, R.~Kearney, H.~Galiana, and
  M.~Verhaegen, ``Subspace identification method for ankle mechanics,'' in
  \emph{Engineering in Medicine and Biology Society, 1995., IEEE 17th Annual
  Conference}, vol.~2, Sep 1995, pp. 1413--1414 vol.2.

\bibitem{MIMO_SS}
V.~Wertz, M.~Gevers, and E.~Hannan, ``The determination of optimum structures
  for the state space representation of multivariate stochastic processes,''
  \emph{Automatic Control, IEEE Transactions on}, vol.~27, no.~6, pp.
  1200--1211, Dec 1982.

\bibitem{MIMO_SS_1}
P.~Stoica and M.~Jansson, ``Mimo system identification: state-space and
  subspace approximations versus transfer function and instrumental
  variables,'' \emph{Signal Processing, IEEE Transactions on}, vol.~48, no.~11,
  pp. 3087--3099, Nov 2000.

\bibitem{MIMO_FIR}
M.~Li and W.~Su, ``Blind identification of mimo fir systems based on multirate
  filterbanks,'' in \emph{Knowledge Acquisition and Modeling Workshop, 2008.
  KAM Workshop 2008. IEEE International Symposium on}, Dec 2008, pp. 936--939.

\bibitem{MIMO_1}
J.~Liang and Z.~Ding, ``Blind mimo system identification based on cumulant
  subspace decomposition,'' \emph{Signal Processing, IEEE Transactions on},
  vol.~51, no.~6, pp. 1457--1468, June 2003.

\bibitem{MIMO_2}
W.~Shi, Q.~Ling, and G.~Wu, ``Sparsity-enhanced linear time-invariant mimo
  system identification,'' in \emph{Control and Decision Conference (CCDC),
  2011 Chinese}, May 2011, pp. 2026--2029.

\bibitem{MIMO_3}
W.~Hachem, F.~Desbouvries, and P.~Loubaton, ``Mimo channel blind identification
  in the presence of spatially correlated noise,'' \emph{Signal Processing,
  IEEE Transactions on}, vol.~50, no.~3, pp. 651--661, Mar 2002.

\bibitem{Ljung_1991}
L.~Ljung, ``Issues in system identification,'' \emph{Control Systems, IEEE},
  vol.~11, no.~1, pp. 25--29, Jan 1991.

\bibitem{given_stat}
B.~Fatimah and S.~D. Joshi, ``Signal matched filter bank with optimized coding
  gain: given statistics case.''

\bibitem{Vaidyanathan}
P.~Vaidyanathan, \emph{Multirate systems and filter banks}.\hskip 1em plus
  0.5em minus 0.4em\relax Eaglewood Cliff, NJ, USA: Prentice-Hall, Inc., 1993.

\bibitem{MIMO_LPTV}
D.~McLernon, ``Relationship between an lptv system and the equivalent lti mimo
  structure,'' \emph{Vision, Image and Signal Processing, IEE Proceedings -},
  vol. 150, no.~3, pp. 133--141, June 2003.

\end{thebibliography}

\end{document}